\documentclass[twocolumn,aps,prl,showpacs]{revtex4}

\usepackage{epsfig}

\begin{document}

\title {Quantum Interference between Impurities:
Creating Novel Many-Body States in $s$-wave Superconductors}
\author{Dirk K.~Morr and Nikolaos A. Stavropoulos}
\affiliation{Department of Physics, University of Illinois at
Chicago, Chicago, IL 60607}
\date{\today}
\begin{abstract}
We demonstrate that quantum interference of electronic waves that
are scattered by multiple magnetic impurities in an $s$-wave
superconductor gives rise to novel bound states. We predict that
by varying the inter-impurity distance or the relative angle
between the impurity spins, the states' quantum numbers, as well
as their distinct frequency and spatial dependencies, can be
altered. Finally, we show that the superconductor can be driven
through multiple local crossovers in which its spin polarization,
$\langle s_z \rangle$, changes between $\langle s_z \rangle=0,
1/2$ and $1$.
\end{abstract}


\maketitle

Over the last two years, several beautiful experiments have
studied quantum interference of electronic waves that are
scattered by multiple impurities \cite{Man00,Der02,Chi02,Hol01}.
In a groundbreaking experiment, Manoharan {\it et
al.}~\cite{Man00} used a corral of magnetic impurities on the
surface of a metallic host to demonstrate that quantum
interference can lead to the focusing of electronic waves into a
quantum image. Moreover, using scanning tunneling spectroscopy
(STS), Derro {\it et al.}~\cite{Der02} were the first to observe
four resonance states in the local density of states (DOS) of the
one-dimensional chains in YBa$_2$Cu$_3$O$_{6+x}$. These states
were interpreted as arising from quantum interference of
electronic waves scattered by two magnetic impurities
\cite{Morr01}. Quantum interference effects were also studied in
optical quantum corrals by Chicanne {\it et al.}~\cite{Chi02} and
between impurities located on quantum dots by Holleitner {\it et
al.}~\cite{Hol01}. Some first theoretical work \cite{theory} has
focused on impurity geometries in metallic systems similar to the
one studied by Manoharan {\it et al.} In contrast, quantum
interference in strongly correlated electron systems, such as
superconductors (with the exception of NbSe$_2$ \cite{Flatte}),
charge- and spin-density-wave systems, or even semi-conductors,
have not yet been addressed. However, the study of interference
effects in theses systems involving spin impurities is not only of
great fundamental interest, but might also possess important
applications in the field of spin electronics \cite{Wolf01} and
quantum information technology \cite{Kane98}.

In order to describe the properties of complex impurity structures
such as quantum corrals, it is first necessary to understand
interference effects associated with the presence of few
impurities. In this Letter we therefore consider two impurities
embedded in a general $s$-wave superconductor (SC). The presence
of two magnetic impurities allows for a coupling of the bound
states associated with a single impurity \cite{Shiba68}, and gives
rise to the emergence of novel many-body states. We show that the
nature of these novel states, i.e., their quantum numbers, can be
altered by varying the distance between the two impurities,
$\Delta r$, or the relative angle between the directions of their
spin moments, $\alpha$. Moreover, we demonstrate that these
changes are accompanied by local crossovers in which the spin
polarization of the superconductor changes between $\langle s_z
\rangle=0, 1/2$, and $1$. We predict that the interplay between
the states' quantum numbers and the inter-impurity distance
determines the distinct frequency and spatial dependence of the
two-impurity bound states. Finally, we discuss the implications of
our work for systems with a larger number of impurities.

Starting point for our calculations is the $\hat{T}$-matrix
formalism \cite{Shiba68} which we generalized to treat the case of
$N$ impurities of spin $S$ with non-magnetic and magnetic
scattering potentials \cite{Morr01,Balatsky}. In the following, we
focus on the case $N=2$, and, following Ref.\cite{Shiba68}, treat
the impurity spins as classical, static variables, corresponding
to the limit $\beta_0=JS/2=const.$~and $S \rightarrow \infty$. In
a fully gaped s-wave SC, this approximation is well justified
since no Kondo-effect occurs for sufficiently small coupling
between the impurities and the delocalized electrons. Within this
approach, any interaction between the impurities is only important
to the extent that it determines the angle, $\alpha$, between the
direction of the impurity spins. Within the Nambu-formalism and in
Matsubara frequency space the electronic Greens function in the
presence of $N$ impurities is given by
\begin{eqnarray}
\hat{G}(r,r',\omega_n)&=&\hat{G}_0(r,r',\omega_n) \nonumber \\
& & \hspace{-2cm} +\sum_{i,j=1}^N
\hat{G}_0(r,r_i,\omega_n)\hat{T}(r_i,r_j,\omega_n)\hat{G}_0(r_j,r',\omega_n)
\ , \label{Ghat}
\end{eqnarray}
where the $\hat{T}$-matrix is obtained from the Bethe-Salpeter
equation
\begin{eqnarray}
\hat{T}(r_i,r_j,\omega_n)&=&\hat{V}_{r_i}\delta_{r_i,r_j}
  \nonumber \\
& & \hspace{-1.5cm} +\hat{V}_{r_i}\sum_{l=1}^N
\hat{G}_0(r_i,r_l,\omega_n)\hat{T}(r_l,r_j,\omega_n) \ .
\label{T1}
\end{eqnarray}
In the case of two impurities
\begin{eqnarray}
& & \hat{V}_{r_1}=\frac{1}{2} \left(U_1 \sigma_0 + J_1S \sigma_3 \right)\tau_3 \ ; \nonumber \\
& & \hat{V}_{r_2}=\frac{1}{2} \left(U_2 \sigma_0 +
J_2S \sigma_3 \cos\alpha+J_2S \sigma_1 \sin\alpha \right)\tau_3 \ ; \nonumber \\
& & \hat{G}^{-1}_0({\bf k},i\omega_n)=\left[ i\omega_n \tau_0 -
\epsilon_{\bf k} \tau_3 \right] \sigma_0 + \Delta_{\bf k} \tau_2
\sigma_2 \ . \label{VG}
\end{eqnarray}
Here, $\hat{V}_{r_1,r_2}$ is the scattering matrix for the
impurities located at ${\bf r}_1$ and ${\bf r}_2$, respectively.
Without loss of generality, we take the spin of impurity $1$ to be
parallel to the $\hat{z}$-axis, while that of impurity $2$ is
rotated from the $\hat{z}$-axis into the $zx$-plane by an angle
$\alpha$. $\hat{G}_0({\bf k},i\omega_n)$ is the Greens function of
the unperturbed (clean) system in momentum space, and $\sigma_i$,
$\tau_i$ are the Pauli-matrices in spin and Nambu-space,
respectively. $U_i$ and $J_i$ are the potential and magnetic
scattering strengths of the impurities. We consider a
two-dimensional (2D) electronic system whose normal state
dispersion is given by $\epsilon_{\bf k}= k^2/2m-\mu$ ($\hbar=1$),
where $\mu=k_F^2/2m$ is the chemical potential, and $k_F=\pi/2$ is
the Fermi wave-vector (we set the lattice constant $a_0=1$). The
results and conclusions presented below are qualitatively robust
against changes in the form of $\epsilon_{\bf k}$, the
dimensionality of the $s$-wave SC, or the size of the
momentum-independent SC gap, $\Delta_{\bf 0}$. For definiteness we
set $\mu=370$ meV and $m^{-1}/\Delta_0=15$, but quantitatively
similar results are obtained for $m^{-1}/\Delta_0=30$. The DOS,
$N({\bf r},\omega)$, presented below is obtained from a numerical
computation of Eqs.(\ref{Ghat})-(\ref{VG}) with $N({\bf
r},\omega)=A_{11}+A_{22}$ and $A_{ii}({\bf r},\omega)=-{\rm Im}\,
\hat{G}_{ii}({\bf r},\omega+i\delta)/ \pi$.

%
%
\begin{figure}[t]
\epsfig{file=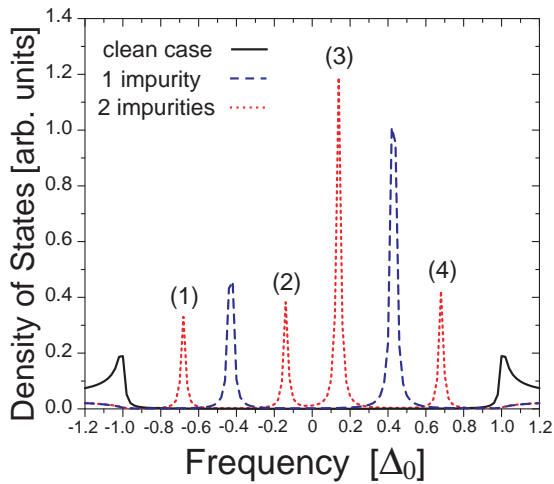,width=7.5cm} \caption{DOS for the clean case
(solid line), a single impurity (dashed line), and two impurities
(dotted line) for $\beta_0=300$ meV (the DOS is shown at the
impurity site).} \label{DOSonsite}
\end{figure}
For a single magnetic impurity in an s-wave SC, the
$\hat{T}$-matrix possesses poles at frequencies
$\omega^{(1,2)}_{res}$, reflecting the presence of two bound
states. The spectroscopic evidence for these bound states are two
peaks in the DOS, as shown in Fig.~\ref{DOSonsite}, where we
present the DOS obtained from Eqs.(\ref{Ghat})-(\ref{VG}) at the
impurity site; for comparison, we also plot the DOS of the clean
system. These results are in general agreement with those of STS
experiments \cite{Yaz97}, which provides further support for the
validity of the $\hat{T}$-matrix approach. Assuming for
definiteness that the impurity spin  $\bf{S} \| \hat{z}$  and
$J>0$, we find that the bound state at $\omega^{(1)}_{res}<0$
($\omega^{(2)}_{res}>0$), which we denoted by $|p,\downarrow
\rangle$ ($|h,\uparrow \rangle$), is particle-like (hole-like)
with spin along the $-z$-direction ($+z$-direction).

We next consider two magnetic impurities with parallel spins,
$U_i=0$ and $J_i=J$. For $\Delta r= \infty $, the two sets of
bound states given by $|p,\downarrow, i \rangle$ and
$|h,\uparrow,i \rangle$ ($i=1,2$) are degenerate. However, for
$\Delta r < \infty$, the probability that an electron scattered by
one of the impurities is also scattered by the second one is
non-zero. Hence, quantum interference of electronic waves that are
scattered by both impurities leads to the formation of novel even
and odd (or bonding and anti-bonding) states, $|p,\downarrow
\rangle_{e,o}= (|p,\downarrow, 1 \rangle \pm |p,\downarrow, 2
\rangle)/\sqrt{2}$, and similarly for the hole-like states. This
picture is confirmed by the numerically computed DOS shown in
Fig.~\ref{DOSonsite} for two impurities located at ${\bf
r}_1=(0,0)$ and ${\bf r}_2=(2,0)$, and $\beta_0=300$ meV (the DOS
shown is that on one of the impurity sites). As expected, the DOS
exhibits four mid-gap peaks with peak $(1),(2)$ corresponding to
the particle-like states $|p,\downarrow \rangle_{e,o}$ and peak
$(3),(4)$ to the hole-like states $|h,\uparrow \rangle_{e,o}$.

\begin{figure}[t]
\epsfig{file=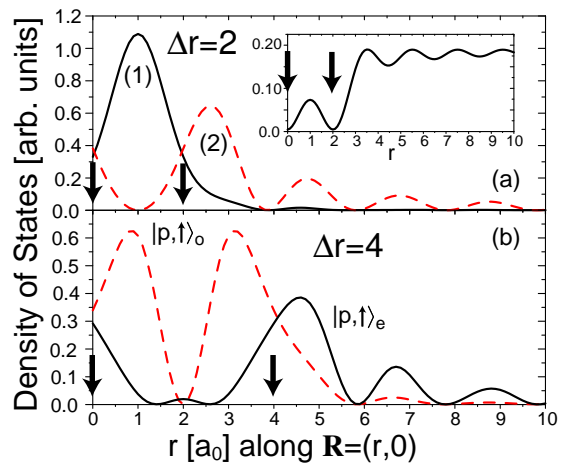,width=7.5cm} \caption{{\it (a)} Spatial
dependence of peak $(1)$ and $(2)$ (see Fig.~\ref{DOSonsite}) in
the DOS along the $\hat{x}$-axis (${\bf R}=(r,0)$) for two
magnetic impurities located at $r=0$ and $r=2$ with parallel
spins. Inset: Spatial dependence of the DOS at
$\omega=\pm\Delta_0$. {\it (b)} Same as {\it (a)} but with
impurities located at $r=0$ and $r=4$.} \label{DOSosc}
\end{figure}

To determine which peaks in the DOS correspond to the even and odd
states, we plot in Fig.~\ref{DOSosc}a the spatial dependence of
the particle-like states $(1)$ and $(2)$ along the $\hat{x}-$axis
with ${\bf R}=(r,0)$ (the location of the impurities at $r=0$ and
$r=2$ are indicated by arrows).
%
%
Since the DOS of the odd states vanishes by symmetry at the
midpoint between the two impurities, i.e., at $r=1$, peak $(2)$
and $(1)$ correspond to the odd and even particle-like states,
respectively. Note that their spatial dependence is remarkably
different: while the odd state exhibits oscillations well beyond
the two impurity region, the even state is primarily confined to
the region between the two impurities. This qualitative difference
is associated with the $(k_F r)$-oscillations of the
$|p,\downarrow ,i\rangle$-states. Since $k_F=\pi/2$ and $\Delta
r=2$, the wave-functions of $|p,\downarrow ,1\rangle$ and
$|p,\downarrow ,2\rangle$ are shifted by a phase $\Delta \phi= k_F
\Delta r=\pi$ outside the two-impurity region ($r>2$ in
Fig.~\ref{DOSosc}a) and their spatial oscillations are
consequently out-of-phase. Thus, $|p,\downarrow ,1\rangle$ and
$|p,\downarrow ,2\rangle$ interfere destructively for
$|p,\downarrow \rangle_{e}$, and only weak spatial oscillations
are observable in the DOS for $r>2$. In contrast, $|p,\downarrow
\rangle_{o}$ shows constructive interference of $|p,\downarrow
,i\rangle$ ($i=1,2$) and its spatial oscillations are enhanced for
$r>2$. By changing the inter-impurity distance to $\Delta r=4$ ,
with $\Delta \phi=2\pi$, the interference pattern between the even
and odd states is exchanged,  and $|p,\downarrow \rangle_{e}$
($|p,\downarrow \rangle_{o}$) now exhibits strong (weak)
oscillations beyond the two impurity region, $r>4$, as shown in
Fig.~\ref{DOSosc}b.

\begin{figure}[t]
\epsfig{file=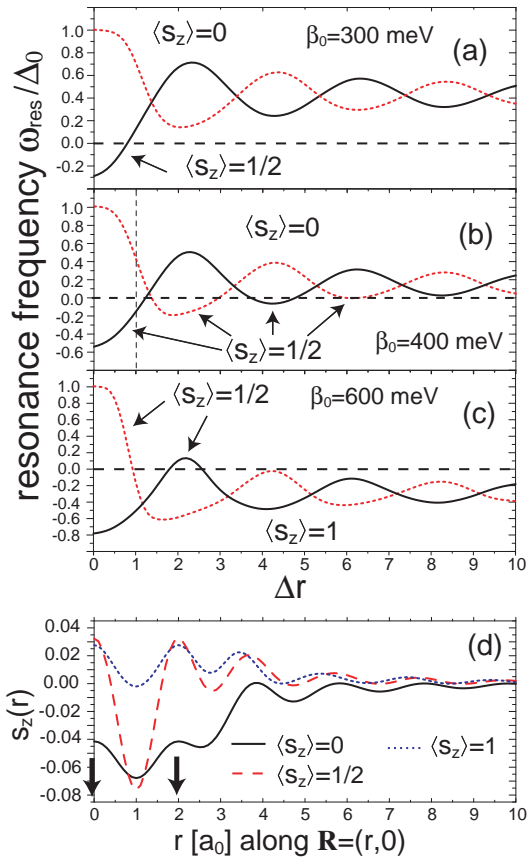,width=7.5cm} \caption{ $\omega_{res}$ for
the even (solid line) and odd state (dotted line) as a function of
$\Delta r$ for parallel impurity spins and (a) $\beta_0=300$ meV;
(b) $\beta_0=400$ meV; and (c) $\beta_0=600$ meV. The ranges of
$\Delta r$ with $\langle s_z \rangle=1/2$ are indicated by arrows.
(d) Local spin polarizations, $s_z(r)$, along the $\hat{x}$-axis,
for $\alpha=0$.} \label{spoles}
\end{figure}

In Fig.~\ref{spoles} we present the bound state energies of the
states $|p,\downarrow \rangle_{e,o}$ as a function of $\Delta r$
for $\alpha=0$ (the frequencies of the $|h,\uparrow
\rangle_{e,o}$-states are obtained via $\omega_{res} \rightarrow
-\omega_{res}$).
%
%
The amplitude of the oscillations in $\omega_{res}$ decays as
$\Delta r^{-d/2} e^{\Delta r/\xi}$ in d-dimensions, but since
$\xi>v_F/\Delta_0 \approx 20$, the exponential decay is barely
perceivable in Fig.~\ref{spoles}. The oscillations' spatial
period, $1/k_F$, directly reflects that of the ($k_F
r$)-oscillations in the wave-functions of $|p,\downarrow
\rangle_{e,o}$, in contrast, to the ($2k_F r$)-oscillations of the
DOS. For certain $\Delta r$, the splitting between the even and
odd states vanishes and the two sets of bound states are
decoupled. This decoupling arises whenever the bound state
wave-functions of one impurity possess a node at the position of
the other impurity. The same type of oscillations were obtained in
Ref.\cite{Flatte} for the case of NbSe$_2$.

At small $\Delta r$, the bound state energy of $|p,\downarrow
\rangle_{e}$ crosses zero, and the state becomes hole-like,
$|h,\downarrow \rangle_{e}$ (at the same time, the state
$|h,\uparrow \rangle_{e}$ transforms into $|p,\uparrow
\rangle_{e}$). We find that this zero-crossing of $\omega_{res}$
is accompanied by a crossover in the spin-polarization of the
superconducting system which at $T=0$ is given by
\begin{equation}
\langle s_z \rangle =\frac{1}{2} \int d^2r \int_{-\infty}^0
d\omega \, \left[ A_{11}({\bf r}, \omega)-A_{22}({\bf r}, \omega)
\right] \ . \label{sz}
\end{equation}
This crossover is similar to the one predicted to occur when the
scattering strength, $\beta_0$, of a single magnetic impurity in
an $s$-wave SC exceeds a critical value, $\beta_c$
\cite{Balatsky,Sak70} (for the band parameters chosen, we obtain
$\beta_c \approx 460$ meV). At this point, the impurity breaks a
Cooper-pair and forms a bound state with one of its electrons.
Specifically, for ${\bf S}\| \hat{z}$ and $J>0$, the spin
polarization changes from $\langle s_z \rangle=0$ for
$\beta<\beta_c$, to $\langle s_z \rangle=1/2$ for $\beta>\beta_c$.
Similarly, for $\beta_0 > \beta_c/2$, the system undergoes a
crossover at $\Delta r_c$, and for $\Delta r< \Delta r_c$ one
electron of the broken-up Cooper-pair forms a single bound state
with {\it both} impurities. For $\Delta r \rightarrow 0$, the DOS
reduces to that of a single magnetic impurity with scattering
strength $2\beta_0$. Accordingly, $|p,\uparrow \rangle_{o}$ moves
towards the particle-hole continuum and vanishes for $\Delta r
\equiv 0$. Finally, a comparison of Fig.~\ref{DOSosc} and
Fig.~\ref{spoles}a shows that, as expected, the spatially more
confined bound state possesses a larger $|\omega_{res}|$ than the
spatially more extended one.

As $\beta_0$ approaches $\beta_c$ from below, the number of
crossovers increases, as shown in Fig.~\ref{spoles}b for
$\beta_0=400 \, {\rm meV} \approx 0.87 \beta_c$. Due to the
oscillatory behavior of $\omega_{res}$, the bound state energy of
the $\downarrow$-state crosses zero for several value of $\Delta
r_{c}$.  As a result, the spin-polarization oscillates between
$\langle s_z \rangle=0$ and $1/2$ and the electronic system can be
tuned through multiple crossovers by varying $\Delta r$. The same
tuning could also be achieved by keeping $\Delta r$ constant and
altering $k_F$ through changes in the doping level using recently
developed field-effect transistor geometries \cite{Sch00}.

For $\beta_0 \geq \beta_c$ the superconductor exhibits a different
crossover in which its spin polarization changes from $\langle s_z
\rangle=1$ to $\langle s_z \rangle=1/2$. For $\Delta r=\infty$,
each impurity breaks one cooper pair and the spin polarization of
the superconducting system is $\langle s_z \rangle=1$. As $\Delta
r$ decreases, one of the bound state energies crosses zero at
least once, as shown in Fig.~\ref{spoles}c where we present
$\omega_{res}$ for the $\downarrow$-states and $\beta_0=600 \,
{\rm meV}\, >\beta_c$. Accordingly, $\langle s_z \rangle$ changes
from $1$ to $1/2$. Since for $\Delta r \rightarrow 0$, the odd
bound state vanishes by symmetry, the spin polarization reaches
$\langle s_z \rangle=1/2$ for any value of $\beta_0 \geq \beta_c$.

Changes in $\langle s_z \rangle$ are also reflected in the
spatially resolved spin polarization, $s_z(r)  = \int_{-\Delta}^0
d\omega (A_{11}-A_{22})$, which due to the limited frequency
integration is experimentally more easily accessible. In
Fig.~\ref{spoles}d, we plot $s_z(r)$ along ${\bf R}=(r,0)$ for two
impurities located at $r=0$ and $r=2$ and $\langle s_z
\rangle=0,1/2,$ and $1$, corresponding to $\beta_0=300, 400$ and
$800$ meV, respectively. For $\langle s_z \rangle=0$, the spin
polarization near the impurities is negative, as expected for
${\bf S} \| \hat{z}$ and $J>0$. For $\langle s_z \rangle=1/2$,
{\it both} impurities form a single bound state with an electron
in the $|\downarrow \rangle_o$-state (see Figs.~\ref{spoles}b).
Thus, $s_z(r)$ is substantially increased at the impurity sites,
but remains practically unchanged at $r=1$. In contrast, for
$\langle s_z \rangle=1$, the electron from the second broken
Cooper-pair joining the two-impurity bound state is in the
$|\downarrow \rangle_e$-state, and consequently, $s_z(r)$
increases primarily around $r=1$. Note, that for two impurities
separated by $\Delta r=4$, the first electron to join the
two-impurity bound state is in the $|\downarrow \rangle_e$-state,
while the second one is in the $|\downarrow \rangle_o$-state, with
corresponding changes in $s_z(r)$.

%
%
\begin{figure}[t]
\epsfig{file=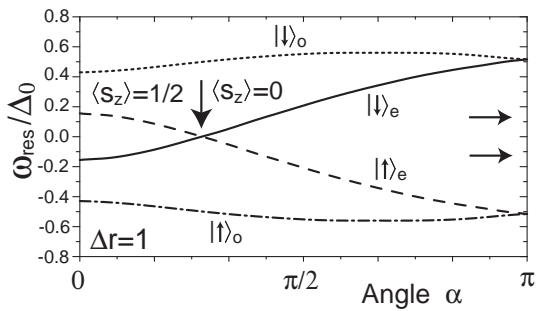,width=7.5cm} \caption{$\omega_{res}$ as a
function of the angle, $\alpha$, between the direction of the
impurity spins for $\Delta r=1$ and $\beta_0=400$ meV. }
\label{Fig4}
\end{figure}

The superconducting system can also be tuned through a crossover
by changing the angle, $\alpha$, between the two impurity spins,
as shown in Fig.~\ref{Fig4}, where we plot $\omega_{res}$ for all
four bound states as a function of $\alpha$ (the impurities are
located at ${\bf r}_1=(0,0)$ and ${\bf r}_2=(1,0)$). Since we
choose $\beta_0=400$ meV $> \beta_c/2$ we have $\langle s_z
\rangle=1/2$ for $\alpha=0$, corresponding to the vertical dashed
line in Fig.~\ref{spoles}b. As $\alpha$ increases from zero, the
frequencies of the even bound states move towards $\omega=0$ which
they cross zero at $\alpha \approx 0.27 \pi$. Simultaneously the
spin polarization changes from $\langle s_z \rangle=1/2$ to
$\langle s_z \rangle=0$. The frequency separation between the even
and odd states of a given spin direction decreases with increasing
$\alpha$ and vanishes at $\alpha=\pi$. This is expected since for
antiparallel impurity spins ($\alpha=\pi$), the bound states for
impurity 1 ($|p,\downarrow ,1\rangle$ and $|h,\uparrow,1 \rangle$)
and impurity 2 ($|p,\uparrow, 2 \rangle$ and $|h,\downarrow, 2
\rangle$) possess different quantum numbers; thus they cannot be
coupled and remain degenerate. However, since the bound states of
one impurity are subjected to the repulsive potential of the
second impurity, their resonance frequencies are larger than those
of a single impurity with the same $\beta_0$ (indicated by the
arrows on the right). This repulsion leads to the disappearance of
all bound states for $\Delta r \rightarrow 0$.

Finally, a non-zero $U$ transfers spectral weight between the
particle- and hole-like states and increases $\beta_c$, but does
not affect our above conclusions. Moreover, a self-consistent
approach that allows for a gap suppression near the magnetic
impurity does not change the qualitative features of the DOS
discussed above \cite{Balatsky}, in agreement with experiment
\cite{Yaz97}.

The results presented above suggest that a superconducting system
with $N$ impurities for which $\beta_0>\beta_c/N$ can be tuned
through multiple crossovers with spin polarizations ranging from
$\langle s_z \rangle=0$ to $N/2$, depending on $\beta_0$, the
inter-impurity distances, and the angles between the spin moments.
Work is currently under way to study these crossovers in more
complex impurity geometries, such as quantum corrals, as well as
the extensions to other host materials, such as unconventional SC,
charge-density-wave systems, or semi-conductors \cite{Morr02b}.

In summary, we show that quantum interference of electronic waves
scattered by two magnetic impurities in an $s$-wave SC gives rise
to novel bound states. We predict that by varying the
inter-impurity distance or the angle between the impurity spins,
the states' quantum numbers can be altered, and the SC can be
driven through multiple local crossovers in which its spin
polarization changes between $\langle s_z \rangle=0, 1/2$ and $1$.

We would like to thank A. Balatsky, J.C. Davis, A. de Lozanne,  M.
Randeria and A. Yazdani for stimulating discussions.

\end{document}